\documentstyle[multicol,aps,epsf]{revtex}

\begin{document}

\title{Condensation of `composite bosons' in a rotating BEC}

\author{N.K. Wilkin $^{(1,2)}$ and J.M.F. Gunn $^{(1,2)}$}

\address{$^{(1)}$ School of Physics and Astronomy, University of Birmingham,
Edgbaston, Birmingham B15 2TT, United Kingdom.}

\address{$^{(2)}$European Synchrotron Radiation Facility, BP. 220, 38043
Grenoble Cedex 9, France}

\date{\today}
\maketitle
\begin{abstract}
We provide evidence for several novel phases in the dilute limit of
rotating BECs. By exact calculation of wavefunctions and energies for
small numbers of particles, we show that the states near integer
angular momentum per particle are best considered condensates of
composite entities, involving vortices and atoms. We are led to this
result by explicit comparison with a description purely in terms of
vortices. Several parallels with the fractional quantum Hall effect
emerge, including the presence of the Pfaffian state.
\end{abstract}


\begin{multicols}{2}
In rotating superfluid $^4$He a vortex lattice forms which, on
scales large compared to the vortex lattice parameter, has a velocity
field indistinguishable from a rigid body co-rotating with the
container. The vortices only perturb the fluid density significantly
over a region of order the coherence length, $\xi$, around the core of
each vortex (of the order of an {\AA}ngstrom). Hence the arrangement of
the vortex lattice is governed by minimising the {\em kinetic} energy
of the fluid in the rotating frame. One may say that the potential
energy is `quenched' by the incompressibility of the fluid.

In the Bose condensed alkali gases although so far it has proved
difficult experimentally to investigate the rotational properties of
the condensates, there has been a vigorous theoretical
debate\cite{fet,rokhsar,Fetterss} about the stability (or otherwise)
of vortices in the condensates. At a mean field level (appropriate for
moderate density), the inhomogeneity of the condensate density and the
existence of surface waves due to the harmonic well makes the
description difficult. Nevertheless the {\em interparticle} potential
energy is still largely unaffected by the presence of vortices in the
limit where the coherence length is small compared to the extent of
the condensate: it is the kinetic energy (and the single-particle
trap potential) which determines the vortex positions.

In this Letter we show that when the coherence length is
comparable to the extent of the condensate, completely new effects
occur. This is due to the {\em kinetic} (and single particle trap) energy being quenched, by a
combination of spherical symmetry and the special properties of the
harmonic well. Hence the ground state in the rotating frame is
determined by the interparticle interactions alone, 
reminiscent of the fractional quantum Hall effect. Indeed we find
stable states that are related to those found in the Hall effect
(albeit in the less familiar regime of filling fraction, $\nu \gtrsim 1$). These include `condensates' of composite bosons of
the atoms attached to an integral number of quanta of angular momenta,
as well as the Laughlin and Pfaffian\cite{MoorRead91} states.

In a rotating reference frame, the standard Hamiltonian for $N$
weakly interacting atoms in a trap is\cite{Stringari_rev}:
\begin{equation}
\vspace*{-3ex}
{\cal H} =\frac{1}{2}\sum_{i=1}^N[-\nabla^2_i + r_i^2 + \eta \sum_{j=1,\ne
i}^N\delta({\mathbf r}_i-{\mathbf r}_j) - 2\bbox{\omega}\cdot {\mathbf
L}_i]
\label{eq:ham}
\end{equation}
where we have used the trap energy, $\hbar \sqrt{K/m} = \hbar\omega_0$
as the unit of energy and the extent, $(\hbar^2/MK)^{1/4}$, of the
harmonic oscillator ground state as unit of length. (Here $M$ is the
mass of an atom and $K$ the spring constant of the harmonic trap.) The
coupling constant is defined as $\eta = 4\pi
{\bar n}a(\hbar^2/MK)^{-1/2}$ where ${\bar n}$ is the average atomic density and $a$ the
scattering length.  The angular velocity of the trap, $\omega$, is
measured in units of the trap frequency.

In the dilute limit $\eta\ll 1$, which implies in
existing experimental traps that the number of atoms, $N$, would be $10 \lesssim N \lesssim 1000$. Then the average coherence (or healing) length is
$\xi\sim 1/\sqrt{{\bar n}a} \to \infty$. It has been shown previously\cite{WGS98} that in
this limit, the problem becomes two-dimensional and the Hilbert space may be truncated to the `lowest Landau
level' states\cite{girvj}, $\psi_m (z)\propto z^me^{-|z|^2/2}$, where
$m\ge 0$ and $z=x+{\rm i}y$ in the plane normal to $\bbox{\omega}$. Indeed at $\omega \!=\!1$ the problem
is identical to the Quantum Hall problem, with $\bbox{\omega}$
replacing the magnetic field.

\begin{figure}[tbp]
\narrowtext
\mbox{
\epsfxsize=\hsize
\epsfbox[140 400  506 720]{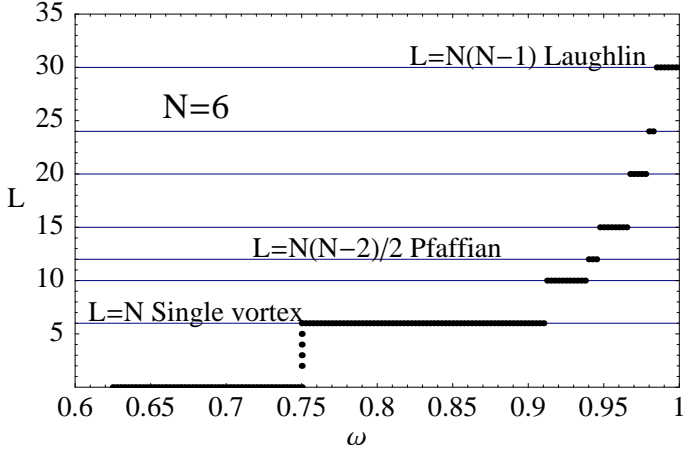}
}
\vspace*{-10ex}
\caption{Stable states for $N\!=\!6$ in the rotating frame with \mbox{$\eta\!=\!1/N$}}
\label{fig:stability}
\end{figure}

We have determined the exact ground state, its energy, $E_0(\omega)$,
and excitation gap, $\Delta$, for the Hamiltonian Eq.~(\ref{eq:ham})
using Mathematica for $N\le 8$ and $\omega \le 1$. In addition we have
determined numerically the lowest eigenvalues for $N\le10$ as a function of
$\omega$. $L_0(\omega)$, the angular momentum of the ground state, is
plotted in Fig. \ref{fig:stability} for $N\!=\! 6$ with $\eta\!=\!1/N$. Angular momentum
remains a good quantum number as we have made no symmetry breaking
ansatz. 

A corresponding plot for $^4$He in a rotating container would show
jumps in the {\em expectation value} of $L_0(\omega)$ as successive
vortices enter the system. The inhomogeneous density of the condensate
in a trap leads to more complex, but similar, behaviour in a mean
field treatment\cite{Rokhn} (appropriate in the high
density limit).  There are a number of important features in
Fig.~\ref{fig:stability}, which are common to all values of $N$ which
we have studied.

 Firstly, at $L\!=\!N$ there is
a state which corresponds\cite{WGS98} to one vortex,
\begin{equation}
\label{eq:ov}
\psi^{\rm 1v}(\{z_i\})=\prod_{i=1}^N (z_i-z_{\rm c}){\rm e}^{-|{\mathbf
z}|^2/2}
\end{equation}
where $z_{\rm c}= (\sum_{i=1}^Nz_i)/N$ is the centre of mass
coordinate and $|{\mathbf z}|^2=\sum_{i=1}^N |z_i|^2$. From this point
we will omit normalisation factors and the ubiquitous ${\rm
e}^{-|{\mathbf z}|^2/2}$.  This state has an interparticle interaction energy
$E=\eta N(N\!-\!2)/4$ and becomes stable at $\omega_1=(1\!-\!N\eta/4)$. (In
addition at $\omega_1$ all $N\ge L\!>\!1$ states are metastable\cite{unpub}.)
Almost all of the other stable states can be labelled by $L\!=\!n(N\!-\!m)$ where $n$
and $m$ are nonnegative integers (this includes the Laughlin state
$n=N$, $m=1$). These values are close to `$n$-vortex states' ($L\!=\!nN$), a possibility we will return to. However the actual wavefunctions for
these states most closely resemble some\cite{JainQ1} used in
the theory of composite fermions\cite{Compf} in the Hall effect. 

We define
$$Q_n (z_i) = \frac{\partial^{(N\!-\!1 \!-\! n)}
}{\partial z_i
{}^{(N\!-\!1 \!-\! n)}} \prod_{j=1, j\ne i}^N (z_i\!-\!z_j)$$
(Note: $\psi^{\rm Lau} = \prod_{i=1}^NQ_{N\!-\!1}(z_i)$ and $\psi^{\rm
1v}= \prod_{i=1}^NQ_1(z_i)$.) Then the states of high overlap with the
true states at $L=n(N\!-\!m)$ may be written as:
$$\psi_{n,m}(\{z_i\}) = \! \!\! \hspace*{-1ex} \sum_{j_1 < j_2 <\cdots < j_{(N\!-\!m)}}^N \!\!\! Q_n (z_{j_1})Q_n(z_{j_2}) \cdots
Q_{n}(z_{j_{(N\!-\!m)}})$$ 
Table 1 shows the overlaps of $\psi_{n,m}$ with the true
ground states for those $L$. Their construction ensures that angular
momentum is used economically to lower the energy: any given particle pair,
$i$ and $j$, will at most be associated with two factors of $(z_i\!-\!z_j)$.

The interpretation of the states at $L\!=\!n(N\!-\!m)$ is that a particle in
association with $n$ quanta of angular momentum is a particularly
stable entity 
in the vicinity of $L\!=\!nN$. As $L$ is reduced $N\!-\!m$ particles remain with all $n$ quanta
and $m$ have all the angular momentum removed. This is our main
result, which occurs at small accessible angular
momenta. This is reminiscent of the `bound state' composite fermions of electrons and vortices\cite{Readsurf}. We will return to this point. 

We will now attempt to reconcile the composite boson states to the vortex states found
in the Nonlinear Schr\"odinger
equation\cite{Fetterss}. The following argument indicates a 
connection. Consider incompressible irrotational fluid (`Helium') in a two-dimensional
circular container, of radius $R$, with $n$ point vortices at radial
coordinates $r_\alpha$. There the angular momentum of the fluid
is\cite{Saff}:
\begin{equation}
\label{eq:ang}
L(\{r_\alpha\}) = N(n\!-\!\sum_{\alpha =
1}^n(r_\alpha/R)^2)
\end{equation}
I.e. the angular momentum is reduced from $L=nN$ by the vortices being
off-centre.

\begin{figure}
\narrowtext
\begin{center}
\mbox{
\epsfxsize=5.5cm
 \epsfbox[91 380 480 717]{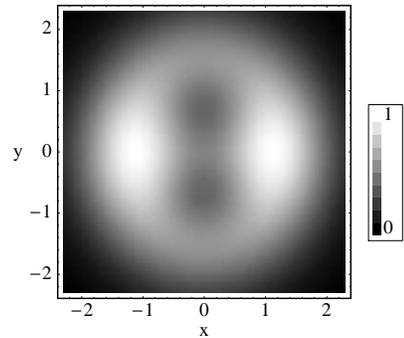}}
\end{center}
\vspace*{-3ex}
\caption{The probability density for the superposition of the states $L\!=\!2N$
and $L\!=\!2N\!-\!2$ for $N\!=\!5$ with `dimples' reminiscent of vortex cores.}
\label{fig:dimple}
\end{figure}

To test this notion, we firstly localise the vortices (resulting in a
non-rotationally invariant state) by 
superposing states with different $L$. Using $L\!=\!10$ and $L\!=\!8$
for $N\!=\!5$ (the Pfaffian state rules out $N=6$) yields the contour plot of probability density,
Fig.\ref{fig:dimple}. The two dimples might be interpreted as two
off-centre vortices (hence the angular momentum is lower than
$L\!=\!2N$). The figure is reminiscent of the figures in \onlinecite{Rokhn}, although the changes in density are
rather small by comparison. Note however, the superposition is
certain to create features periodic with $\cos 2\theta$,
where $\theta$ is the polar angle. 
To quantify these ideas, we introduce the lowest Landau level vortex factors, $\prod_{i=1}^N
(\zeta_\alpha - z_i)$ with complex vortex  coordinate
$\zeta_\alpha$. A complete set of {\em particle} states of angular
momentum $L$ is obtained using $n\!=\!L$ vortices: 
$$\psi(\{z_i\}) = \int \prod_{\beta=1}^n {\rm
d}^2\zeta_\beta{\rm e}^{-|\bbox{\zeta}|^2} \phi(\{\zeta_\alpha\})
\prod_{\beta=1}^n\prod_{i=1}^N (\zeta_\beta - z_i) 
 $$
where $|\bbox{\zeta}|^2= \sum_{\alpha=1}^n |\zeta_\alpha|^2$. This
follows from the result \cite{Mehta} that an $n$th order
polynomial in the $z$'s may be expanded using the products of
powers of the elementary symmetric polynomials, $C_r$, $0\le r\le n$: 
$$C_r(\{z_i\}) =
\sum_{i_1<i_2<\cdots<i_r} z_{i_1}z_{i_2}\cdots z_{i_r}$$ 
Although there is no need to write wavefunctions of {\em both} the
particle, $z_i$, and the vortex coordinates, $\zeta_\alpha$, it will be convenient. 

It will be useful to note the form of the resulting particle wave
function, $\psi$, when there is one vortex. The construction implies the natural one-vortex
states are $\phi_p(\zeta^*) = \zeta^{*p}/(p!\pi)$. The corresponding
particle states are $\psi_{N\!-\!p}
\propto C_p(\{z_i\})$ for $0\le p \le N$; and for $p>N$
$\psi_{N\!-\!p} =0$. Note that the angular momentum of the particles is $L=N\!-\!p$, consistent
with the special cases: $\psi_N = 1$
i.e. placing a vortex in this state has no effect on the non-rotating condensate;
$\psi_{N\!-\!1} = \sum_{i=1}^N z_i = z_{\rm c}$; and $\psi_0 = \prod_{i=1}^Nz_i$, i.e. a `simple' single vortex.

It is tempting to relate the stable states at $L=n(N-m)$, to
Eq.~\ref{eq:ang}, as being $n$
vortices in state $p=m$, with associated displacement of the vortex
determined by $\langle
|\zeta|^2\rangle_p = p+1$. However this implies $r_{\rm v}^2= p+1$ and
this leads to a contradiction unless $L\sim N^2$ (using the empirical
relation $p=m\le n$), which is too restrictive.  

Moreover, this purely vortex description, $\phi(\zeta)$, requires more vortices than $n$ (in
$L=n(N\!-\!m)$). For example $L\!=\!N$: 
expanding the product in Eq.\ref{eq:ov} we see there is a term $z_{\rm c}^N = C_1(\{z_i\})^N$ whose generation requires $N$ vortex
factors (even more for larger $L$). In addition, the
number of vortices is not fixed, as the number in the vortex state
$\zeta^N$ is indeterminate since they do not affect the particle
wave-function (in a sense it is the vortex vacuum state). This is in stark contrast to the
incompressible ($\xi \to 0$) case where the number of vortices is
fixed and they are classical entities.

\begin{figure}
\narrowtext
\mbox{
\epsfxsize=5cm
\epsfbox[91 502 439 720]{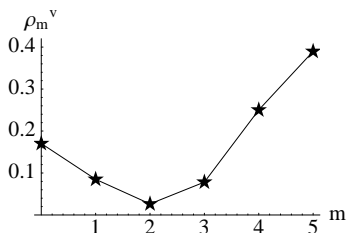}
}
\caption{Eigenvalues ($\rho^{\rm v}_m$) of the single {\em vortex} density matrix
 for eigenfunctions $\zeta^m$ for $L\!=\!2N\!-\!2$ for $N\!=\!6$. The
trace is normalised to unity, having suppressed the
weight associated with the vortex `vacuum' state, $m=N$.}
\label{fig:condv}
\end{figure}

It might be supposed that although there may be a fluctuating vortex
population at large vortex quantum numbers, this is in the tail of the
particle wavefunction and the description may be simple near the centre
of the trap. This is determined by computing the single-{\em vortex} density
matrix, $\rho^{\rm v}(\zeta,\zeta')$, for the exact state $L\!=\!2N\!-\!2$,
$N\!=\!6$, Fig.~\ref{fig:condv}.
If the displaced vortex picture were correct, one might expect a
factor in the vortex wavefunction of the form
$(\zeta^*_1-\zeta^*_2)^2$ corresponding to the vortices rotating
around the centre of the trap. This
factor alone would lead to the following eigenvalues, $\rho_m^{\rm v}$
(with corresponding eigenvectors $\zeta^m$),
of $\rho^{\rm v}$: $\rho^{\rm
v}_0=\frac{1}{4}$, $\rho^{\rm v}_1=\frac{1}{2}$ and $\rho^{\rm v}_2=
\frac{1}{4}$. As can be seen from Fig.\ref{fig:condv}, this is not the
case. The most pronounced feature is a {\em maximum} at $m\!=\!0$. The
vortices tend to condense, in the $m\!=\!0$ state, not to separate in
$|\zeta^*_1-\zeta^*_2|$. (Further evidence comes from evaluating the particle density matrix \cite{unpub}).

These difficulties in describing the $\psi_{n,m}$
states purely in vortex variables occur because the particles are
{\em binding} to the vortices. This leads to a strongly correlated
state whose description requires additional vortex variables if they
are used alone. One interpretation uses 
ideas from the quantum Hall effect\cite{Readsurf}: At the centre of each
vortex there is a decrease in the particle density. Thus in terms of
interparticle interactions, this is a low energy region for an
additional particle. 

Mathematically this is described most easily for the Laughlin state,
using $N$ vortices, $\zeta_\alpha$, with a factor $\prod_{i, i\ne
\alpha}^N(\zeta_\alpha - z_i)$ where the $\alpha$-th particle experiences
no suppression of its amplitude: it is `bound'. This can be expressed as 
$$\psi^{\rm Lau} (\{z_i\}) = \int \prod_{\beta=1}^N {\rm
d}^2\zeta_\beta{\rm e}^{-|\bbox{\zeta}|^2}
{\rm e}^{{\mathbf
z}\cdot{\bbox \zeta}^*} \prod_{\alpha \ne j}(\zeta_\alpha - z_j) $$
(Noting ${\rm e}^{ z\eta^* }$ and $(-1)^n \eta^{*n}{\rm e}^{z\eta^*}$
respectively play the roles\cite{barg,girvj} of a delta
function and its $n$th derivative within the lowest Landau level.) I.e.
the delta function factors bind the $i$th particle to
the $i$th vortex. 

To generate the states $\psi_{n,m}$ we use the derivatives of the
Lowest Landau level delta function so that:
$$\psi_{n,m} (\{z_i\}) = \int \prod_{\beta=1}^N {\rm
d}^2\zeta_\beta{\rm e}^{-|\bbox{\zeta}|^2}
{\rm e}^{{\mathbf
z}\cdot{\bbox \zeta}^*} \phi_{n,m}(\{\zeta^*_\gamma\}) \prod_{\alpha \ne
j}(\zeta_\alpha - z_j)$$
where
$$\phi_{n,m}(\{\zeta_\alpha\}) = \prod_{\beta = 1}^N \zeta_\beta^{*(N-1-n)}
\sum_{\gamma_1<\gamma_2<\cdots <\gamma_m}^N
\zeta_{\gamma_1}^{*n}\zeta_{\gamma_2}^{*n}\cdots
\zeta_{\gamma_m}^{*n}$$
which can be interpreted as a `condensate' of $(N\!-\!m)$ composites (each
consisting of an atom
and a vortex) in the state, $\zeta^{*r}$ with $r=(N\!-\!1\!-\!n)$. The
remaining unbound atoms remain condensed in the single particle ground
state. (The states $L=n(N-m)$ are also selected using a composite
fermion approach \cite{nrc}.)

The remaining stable states are consistent with the bosonic Pfaffian
state\cite{MoorRead91,Read98nu1}, at $L= \frac{1}{2} N(N\!-\!2)$ for even
$N$ and $L=\frac{1}{2} (N\!-\!1)^2$ for odd $N$. 
$$
\psi^{\rm Pf}(\{z_i\}) = \prod_{i< j} (z_i-z_j) {\rm
 Pf}\left(\frac{1}{z_i-z_j}\right)
$$
where the Pfaffian is defined
$$
{\rm Pf} \left(\frac{1}{z_i-z_j}\right)={\cal A}
\left[\frac{1}{(z_1-z_2)}\frac{1}{(z_3-z_4)}\cdots
\frac{1}{(z_{N-1}-z_N)}\right]
$$
where $\cal A$ denotes antisymmetrisation of the following
product. (This is generalised for odd $N$ by omitting one of the
particles in each term of the antisymmetrisation
\onlinecite{Mehta}.) The overlaps of $\psi^{\rm Pf}$ with the exact
ground state for $N\!=\! 5$ and $L\!=\! 8$, $N\!=\! 6$ and $L\!=\! 12$ and $N\!=\!7$ and
$L\!=\!18$ are: $0.91^2$, $0.90^2$ and $0.80^2$.

Some nearby stable states, e.g. $L\!=\!10$ and $L\!=\!14$,
are well described as simple modifications of the Pfaffian state. This uses the conjecture (which has been demonstrated by direct
evaluation for $4\!\le\!N\!\le\!8$) that the Pfaffian state may be
represented by a product of two Laughlin states for $N/2$ particles
(or for odd $N$, a cluster of $(N\!-\!1)/2$ and one of $(N+1)/2$):
$$\psi^{\rm Pf}(\{z_i\}) = {\cal S}  \prod_{i<j \in \sigma_1}(z_i-z_j)^2  \prod_{k<l\in 
\sigma_2}(z_k-z_l)^2$$
where the two subsets, $\sigma_1$ and $\sigma_2$, each have $N/2$
particles ($(N\!-\!1)/2$ and $(N+1)/2$ for odd $N$). $\cal S$
indicates that the wave function is symmetrised over the distribution
of the particles into these subsets. These two  well-correlated clusters
appear to be `dual' to the clusters of Halperin\cite{Halp}
which have a high internal energy, due to the lack of nodal factors. 

For example the state $N\!=\!6$, $L\!=\!14$ has overlap $0.96^2$ with a state
with two quanta of angular momenta in the centre of mass motion of the
clusters (defining $Z_b= \sum_{i \in \sigma_b}z_i$, $b=1$ or $2$):
$$\psi^{L\!=\!14} (\{z_i\}) ={\cal S}(Z_1 - Z_2)^2\psi^{\rm Pf}$$
The state  $N\!=\!6$, $L\!=\!10$ has overlap $0.97^2$ with a state where there is one factor of centre of mass motion and one vortex has been `removed' from one of the clusters:
$$\psi^{L=10} (\{z_i\}) ={\cal S}(Z_1 - Z_2) \psi^{\rm Pf}\prod_{p \in
\sigma_1}\sum_{q< p, q\in \sigma_1}\frac{1}{z_p-z_q}$$
(The apparent asymmetry of the last factor only involving the first cluster, $\sigma_1$, is illusory due to the overall symmetrisation.)

In conclusion, this Letter provides evidence that the weak coupling
limit of rotating BEC's contains some novel phenomena. These occur
even in the regime where one would anticipate small numbers of
vortices and hence should be open to experimental investigation in the
near future. 

\acknowledgements{We would like to thank J.T. Chalker, N.R. Cooper,
D.S. Rokhsar and R. A. Smith for helpful discussions. We are grateful
to P. Carra and the ESRF for hospitality whilst this work was
completed and ITP (PHY-94-07194) during the early stages of the
work. We acknowledge financial support from the EPSRC
GR/L28784,GR/K6835 and GR/L29156.}

\end{multicols}
\begin{multicols}{1}
\begin{minipage}[h]{18cm}
\begin{center}
\footnotesize
\begin{tabular}
{|p{0.5cm}|p{0.82cm}|p{0.82cm}|p{0.82cm}|p{0.82cm}|p{0.82cm}|p{0.82cm}|p{0.82cm}|p{0.82cm}|p{0.82cm}|p{0.82cm}|p{0.82cm}|p{0.82cm}|p{0.82cm}|p{0.82cm}|}\hline
\mbox{  }&{\scriptsize $2N\!-\!4$}&{\scriptsize $2N\!-\!2
$}&{\scriptsize $2 N \hspace*{4ex}$}&{\scriptsize $3N\!-\!6 $}&{\scriptsize $3N\!-\!3$}&{\scriptsize $3N \hspace*{4ex}$} &{\scriptsize $4N\!-\!8 $}&{\scriptsize $4N\!-\!4 $}&{\scriptsize $4N \hspace*{4ex}$}&{\scriptsize $5N\!-\!10$}&{\scriptsize $5N\!-\!5 $}&{\scriptsize $5N \hspace*{4ex}$}&{\scriptsize $6N\!-\!6 $}&{\scriptsize $6N \hspace*{5ex}$}\\\hline\hline
5     & &{\bf 	8}\mbox{ } $\spadesuit$ $0.98^2$ &
{\bf   10 } $0.87^2$  &	     & {\bf 12 } $0.99^2$&{\bf  15} $0.96^2$&	&      &  {\bf 	20} \newline 1  &&    &    &      &\\[1ex]\hline
6 &     &{\bf  10 }  $\spadesuit$ $0.86^2$ & {\bf  12} \mbox{ }$\spadesuit$ $0.69^2$ & {\bf 12}\mbox{ }$\spadesuit$  $0.79^2$  &
{\bf 15} $0.95^2$ &     &      &{\bf 20} $0.99^2$   &    {\bf
24} $0.94^2$ & &      &  {\bf 30}  \newline 1 &      &\\[1ex]\hline
7 &       & {\bf 12 } $0.83^2$ &      & {\bf 15}  $0.49^2$  & {\bf
18}\mbox{ }$\spadesuit$ $0.85^2$ &      &      &{\bf 24 }   &      &    &{\bf 30 }  & {\bf35 }&
&{\bf42}  \newline 1 \\[1ex]\hline
8 &    {\bf  12} $0.66^2$ & {\bf14}   &      &{\bf 18  }  &      &{\bf
24}\mbox{ }$\spadesuit$ &{\bf24 } \mbox{ }$\spadesuit$  &
&{\bf 32}& {\bf 30} & {\bf35}   &    & {\bf  42} & $\ldots$\\[1ex]\hline\hline
\end{tabular}
\normalsize

\vspace{2ex}\end{center}
{\small TABLE 1. Stable states for $N \le 8$: the upper number
is their angular momentum and the lower is their overlap with the $Q$
wavefunctions. $\spadesuit$
indicates that the wavefunction can also be written (or derived from) a Pfaffian state.}
\end{minipage}
\end{multicols}
\vspace*{-0.1cm}

\begin{minipage}{18cm}
\begin{multicols}{2}

\end{multicols}
\end{minipage}


\begin{thebibliography}{10}
\vspace*{-0.6cm}
\bibitem{fet}
A.~A. Svidzinsky and A.~L. Fetter, preprint, cond-mat/9811348.

\bibitem{rokhsar}
D.~S. Rokhsar, Phys.\ Rev.\ Lett. {\bf 79},  2164  (1997).

\bibitem{Fetterss}
A.~L. Fetter, lectures from \mbox{``Enrico Fermi''} School on BEC,  preprint cond-mat/9811366.

\bibitem{MoorRead91}
G. Moore and N. Read, Nucl.\ Phys.\ B {\bf 360},  362  (1991).

\bibitem{Stringari_rev}
F. Dalfovo, S. Giorgini, L.~P. Pitaevskii, and S. Stringari, Rev.\ Mod.\ Phys.
  {\bf 71},  463  (1999).

\bibitem{WGS98}
N.~K. Wilkin, J.~M.~F. Gunn, and R.~A. Smith, Phys.\ Rev.\ Lett. {\bf 80},
  2265  (1998).

\bibitem{girvj}
S.~M. Girvin and T. Jach, Phys.\ Rev.\ B {\bf 28},  4506  (1983).

\bibitem{Rokhn}
D.~A. Butts and D.~S. Rokhsar, Nature {\bf 397},  327  (1999).
\newpage
\bibitem{unpub}
N.~K. Wilkin, J.~M.~F. Gunn, and R.~A. Smith, \mbox{unpublished results.}

\bibitem{JainQ1}
G. Dev and J.~K. Jain, Phys.\ Rev.\ B {\bf 45},  1223  (1992).

\bibitem{Compf}
\mbox{edited by O. Heinonen}, {\em Composite fermions: a unified view of the
  quantum hall regime} (World Scientific, Publishing Co., 1998).

\bibitem{Readsurf}
N. Read, Surf. Sci. {\bf 362},  7  (1996).


\bibitem{Saff}
P.~G. Saffman, {\em Vortex dynamics} (Cambridge University Press, Cambridge,
  1995).

\bibitem{Mehta}
M.~L. Mehta, {\em Matrix Theory: Selected Topics and Useful Results} (Les
  Editions de Physique, Les Ulis, 1989), p. 254, p. 15.

\bibitem{barg}
V. Bargmann, Rev. Mod. Phys. {\bf 34},  829  (1962).

\bibitem{nrc}
N. R. Cooper, private communication.

\bibitem{Read98nu1}
N. Read, Phys.\ Rev.\ B {\bf 58},  16262  (1998).

\bibitem{Halp}
B.~I. Halperin, Helv. Phys. Acta {\bf 56},  75  (1983).

\end{thebibliography}
\end{document}